\documentclass[conference]{IEEEtran}
\IEEEoverridecommandlockouts
\usepackage{cite}
\usepackage{amsmath,amssymb,amsfonts}
\usepackage{algorithmic}
\usepackage{graphicx}
\usepackage{textcomp}
\usepackage{colortbl}
\usepackage[most]{tcolorbox}
\usepackage{balance}
\usepackage{orcidlink}

\tcbset{textmarker/.style={
        enhanced,
        parbox=false,boxrule=0mm,boxsep=0mm,arc=0mm,
        outer arc=0mm,left=5mm,right=3mm,top=7pt,bottom=7pt,
        toptitle=1mm,bottomtitle=1mm}}

\newtcolorbox{noteBox}{textmarker,
    borderline west={4pt}{0pt}{gray},
    colback=gray!10!white}

\newcommand{\note}[1]{\begin{noteBox} #1 \end{noteBox}}

    \usepackage{xcolor}
\def\BibTeX{{\rm B\kern-.05em{\sc i\kern-.025em b}\kern-.08em
    T\kern-.1667em\lower.7ex\hbox{E}\kern-.125emX}}
\begin{document}

\title{Beyond the Code: A Multi-Modal Assessment Strategy for Fostering Professional Competencies via Introductory Programming Projects}

\author{\IEEEauthorblockN{Santiago Berrezueta-Guzman\,\orcidlink{0000-0001-5559-2056}}
\IEEEauthorblockA{
\textit{Technical University of Munich}\\
Heilbronn, Germany \\
}
\and
\IEEEauthorblockN{Vanesa Metaj\,\orcidlink{0009-0008-6576-346X}}
\IEEEauthorblockA{
\textit{Technical University of Munich}\\
Heilbronn, Germany \\
}
\and
\IEEEauthorblockN{Stefan Wagner\,\orcidlink{0000-0002-5256-8429}}
\IEEEauthorblockA{
\textit{Technical University of Munich}\\
Heilbronn, Germany \\
}
}

\maketitle

\begin{abstract}
As the landscape of software engineering evolves, introductory programming courses must go beyond teaching syntax to foster comprehensive technical competencies and professional soft skills. This paper reports on a pedagogical experience in a "Fundamentals of Programming" course that used a Project-Based Learning (PBL) framework to develop a 2D "Maze Runner"-style game. While game development serves as a high-engagement vehicle for mastering core concepts, such as multidimensional arrays, control structures, and logic, the core of this study focuses on implementing a rigorous, multifaceted assessment model structured across four distinct dimensions: (1) an in-situ technical demonstration, evaluating real-time code execution and algorithmic robustness; (2) a technical screencast, requiring students to articulate their work in a concise audiovisual format; (3) a formal presentation to instructors, defending their project's design patterns and problem-solving strategies; and (4) a structured peer-review process, where students evaluated their colleagues' projects.

Our findings suggest that this multi-dimensional approach not only improves student retention of programming fundamentals but also significantly enhances communication skills and critical thinking. By integrating peer evaluation and multimedia documentation, the course successfully bridges the gap between basic coding and the collaborative requirements of modern software engineering. This paper details the curriculum design, the challenges of implementing diverse assessment pillars, and the measurable impact on student performance and engagement, providing a scalable roadmap for educators looking to modernize introductory computing curricula.
\end{abstract}


\section{Introduction}\label{I}

Software engineering is no longer just about writing code that works. The field has evolved into a discipline that demands collaboration, communication, documentation, and the ability to reason about design decisions under real constraints \cite{ouhbi2020software, tembrevilla2024experiential}. Yet introductory programming courses have been slow to reflect this shift \cite{medeiros2018systematic}. 

The dominant model still centers on syntax, small exercises, and individual assessments, all of which are easy to grade but not well-suited to measure whether a student can actually function on a software development team \cite{rehman2023trends}. The rapid leverage of AI coding assistants has only increased the need for better assessment. Tools like GitHub Copilot or ChatGPT can now generate perfect solutions to programming tasks, making it very difficult to rely on functional correctness alone as a requirement for student competence \cite{berrezueta2025coders, becker2023programming}. As a result, educators are under growing pressure to rethink not just what they teach, but also how they assess it \cite{berrezueta2023recommendations}.

This pressure is especially acute in introductory courses, where most students arrive with no prior programming experience and are expected to build these skills during the semester \cite{rodrigues2022introductory}. The most valued skills in the industry, such as clear communication, structured thinking, peer collaboration, and the ability to explain one's own code, are precisely the ones that traditional assessment methods are least able to capture \cite{messer2024automated}. A student who can pass a written test on object-oriented principles may still struggle to document their decisions or justify their implementation. The gap between mastering syntax and professional competency will continue to grow the longer it goes unaddressed \cite{garousi2019closing}. Correctly addressing this issue requires assessment models that go beyond the code itself, treating the project not just as a deliverable but as a way to develop and evaluate a broader set of competencies \cite{che2025formative}.

We propose and evaluate a four-dimensional assessment framework for introductory programming courses that simultaneously targets technical competency and professional soft skills. We demonstrate how a game development project can serve as a vehicle for integrating in-situ mentoring, multimedia documentation, live defense, and structured peer evaluation within a single cohesive model. We further provide empirical evidence, drawn from a cohort of 138 first-semester students, on the effectiveness and feasibility of this approach, and offer a scalable roadmap for educators seeking to modernize introductory computing curricula beyond functional code assessment.

The remainder of this paper is organized as follows. Section~\ref{RW} reviews related work. Section \ref{CCD} describes the course context, curriculum design, and the project framework. Section~\ref{M} details each dimension of the multi-modal assessment model. Section~\ref{IC} discusses the implementation challenges during deployment. Section~\ref{R} presents the results and analysis. Section~\ref{D} discusses the broader implications and limitations of the study. Finally, Section~\ref{C} summarizes the conclusions and outlines directions for future research.

\section{Related Work}\label{RW}

Prior work in this area spans four interconnected domains: Project-Based Learning in computing education, game development as a pedagogical tool, assessment models in introductory programming courses, and peer review in collaborative learning environments.

Rehman \cite{rehman2023trends} conducted a systematic literature review to analyze the evolution and implementation difficulties of Project-Based Learning in Computer Science Education. The study identifies assessment as a primary challenge, noting that traditional testing methods fail to capture skills, such as teamwork or communication, which are developed during PBL. It is also emphasized that educators have been adopting new evaluation techniques, such as peer reviews and self-assessments, to accurately track students' learning progress and provide better feedback.

Garcia et al. \cite{garcia2019game} explored the effectiveness of game development as a pedagogical tool using the Construct 2 engine, finding that the creative process significantly improved learner engagement and perceived skill acquisition in multimedia and logic. 
Johnson et al. \cite{10.1145/3024906.3024908} also agree that such approaches foster active learning and motivation. Still, their systematic review of over 100 educational games found a significant lack of empirical evidence supporting these outcomes. They argue that, although the potential of game-based learning is high, the community currently lacks sufficient assessment frameworks to rigorously measure technical competency and learning outcomes.

Messer et al. \cite{messer2024automated} found that while Automated Assessment Tools are highly effective for grading functional correctness via dynamic unit testing, they often fail to assess code quality attributes such as maintainability, readability, and documentation. To address these limitations, it is necessary to diversify assessment strategies. 

Che Abdul Rani et al. \cite{che2025formative} reviewed formative assessment methods in higher education and found that interactive strategies, particularly peer assessment and games, significantly enhance student engagement and critical thinking. Their findings suggest that a diverse mix of assessment techniques is superior to a single one, as it better aligns with the different qualifications of programmers.

Lin et al. \cite{https://doi.org/10.1002/cae.22425} conducted a randomized control trial to investigate how peer code review (PCR) influences computational thinking and student engagement in blended learning environments. Their findings showed that students participating in structured peer reviews demonstrated significantly higher performance in computational concepts and perspectives, as well as increased learning satisfaction. Rathna Sekhar and Swapna Goud \cite{article} complemented this by exploring a broader range of collaborative teaching and learning methods in Python programming courses. Their case study showed that the experimental group using these methods achieved significantly higher average marks and better knowledge retention than a group taught through traditional methods.

As summarized in Table~\ref{tab:related_work}, while prior work validates individual strategies such as game-based engagement \cite{garcia2019game, 10.1145/3024906.3024908}, automated grading \cite{messer2024automated}, and peer review \cite{https://doi.org/10.1002/cae.22425, article, che2025formative}, no study embeds all of these within a single cohesive framework targeting first-semester students. This work addresses that gap by proposing and evaluating a four-dimensional assessment model that simultaneously develops technical and professional competencies.

\begin{table}[h]
\centering
\caption{Comparison of Related Work Against the Proposed Framework}
\label{tab:related_work}
\begin{tabular}{|p{1.6cm}|p{0.6cm}|p{0.6cm}|p{0.9cm}|p{1.6cm}|p{0.8cm}|}
\hline
\textbf{Study} & \textbf{PBL} & \textbf{Game Dev.} & \textbf{Peer Review} & \textbf{Multi-Modal Assessment} & \textbf{Intro. Course} \\
\hline
Rehman \cite{rehman2023trends} & \checkmark & & \checkmark & & \\
\hline
Garcia et al. \cite{garcia2019game} & & \checkmark & & & \\
\hline
Johnson et al. \cite{10.1145/3024906.3024908} & & \checkmark & & & \\
\hline
Messer et al. \cite{messer2024automated} & & & & & \checkmark \\
\hline
Che Abdul Rani et al. \cite{che2025formative} & \checkmark & & \checkmark & & \\
\hline
Lin et al. \cite{https://doi.org/10.1002/cae.22425} & & & \checkmark & & \\
\hline
Rathna Sekhar et al. \cite{article} & & & \checkmark & & \checkmark \\
\hline
\rowcolor{lightgray} \textbf{This work} & \checkmark & \checkmark & \checkmark & \checkmark & \checkmark \\
\hline
\end{tabular}
\end{table}

\section{Course and Curriculum Design}\label{CCD}

\subsection{Course Context and Learning Objectives}
\textit{Fundamentals of Programming (FoP)} is a compulsory first-semester course within the B.Sc of Information Engineering. It is designed to be accessible to students with no prior programming background, and it is offered concurrently with the \textit{Introduction to Informatics} course, bridging theoretical concepts with hands-on implementation. The syllabus follows a structured progression, beginning with control structures, data types, and algorithmic reasoning, and advancing toward object-oriented programming principles. The curriculum further encompasses graphical user interface development, recursion, and programming paradigms beyond Java, with the explicit goal of aligning student competencies with contemporary software engineering practices. By the end of the course, students are expected to independently develop correct Java programs while demonstrating the capacity to apply systematic problem-solving strategies and object-oriented design principles to real-world scenarios.

\subsection{Student Demographics and Prerequisites}
The course consisted of 138 students, approximately 70\% male and 30\% female. Students represented 31 nationalities, with the largest groups originating from East and South Asia. The official communication language was English. 

No prior programming experience was required for enrollment. The course was designed on the assumption that all relevant skills would be acquired from the ground up during the semester. A small number of students arrived with some self-taught exposure and initially progressed at a noticeably faster pace. However, this advantage narrowed as the semester progressed, because many students managed to catch up on time and practice.

\subsection{The PBL Framework: The ``Maze Runner'' Game Project}
\textit{FoP} requires students not only to solve homework and in-class exercises, but also to complete a project. Organized in teams of three, the project accounts for 60\% of the final grade, split between development, screencast, peer review, and presentation.

The project tasks students with building a 2D maze game in which a character navigates through a maze, collects keys to unlock exits, avoids or confronts obstacles, and escapes before losing all lives. Beyond the technical requirements, students have to develop an original background story and align their visual, audio, and design choices around it, making creativity another grading criterion. This encourages students to go beyond simply fulfilling programming requirements.

Teams are provided a project skeleton, but receive no formal instruction in project coordination, task distribution, or AI tool usage, leaving the organizational side entirely up to them. Minimum requirements cover functional completeness, code structure, and documentation. Teams that go beyond this baseline with creative extensions to the game are rewarded with bonus points, keeping the project accessible to all students while giving more motivated teams room to push further.

\subsection{Core Programming Concepts Addressed}
The Maze Runner project is designed to serve as a task that requires students to apply every concept introduced throughout the semester in a functional game. At the most foundational level, students must work fluently with control structures and data types to handle game logic such as collision detection, life management, and key collection. The maze itself is loaded from a Java properties file, which demands a solid understanding of data structures and file I/O before any game logic can be built.

Object-oriented design is the most heavily exercised competency. Students are required to model each game element as dedicated classes sharing a common superclass hierarchy, applying inheritance, encapsulation, and polymorphism.
Algorithms are also central to pathfinding within a certain range of the player. Camera behavior, viewport scaling, and maze validation demand careful reasoning about coordinate systems and spatial logic. Recursion and more advanced data structures, introduced later in the semester, naturally find application in maze traversal or game state management.

\subsection{Project Milestones and Timeline}

The project followed a structured timeline. It was announced in the fourth week of the semester, giving students roughly ten weeks to work on it alongside their regular homework and in-class exercises.
Beyond the development itself, students had two additional project-related tasks. Each team had to complete a peer review, anonymously evaluating two other projects using the same grading requirements as the instructors. They also had to submit a screencast, which is a short recorded video in which they demonstrate their game and explain key technical decisions. Both tasks were completed after the project submission, in the final weeks before presentations.

Teams that met the minimum requirements were then invited to present their project in person to instructors in a 20-minute session. They had to cover a live demo followed by a Q\&A. The three best screencasts and the three best projects were exempt from this and received full presentation points automatically.
Together, these components form the multi-modal assessment model that is the focus of this paper.

\section{The Multi-Dimensional Assessment Model}\label{M}

\begin{figure*}[h!]
    \centering
    \includegraphics[width=1\linewidth]{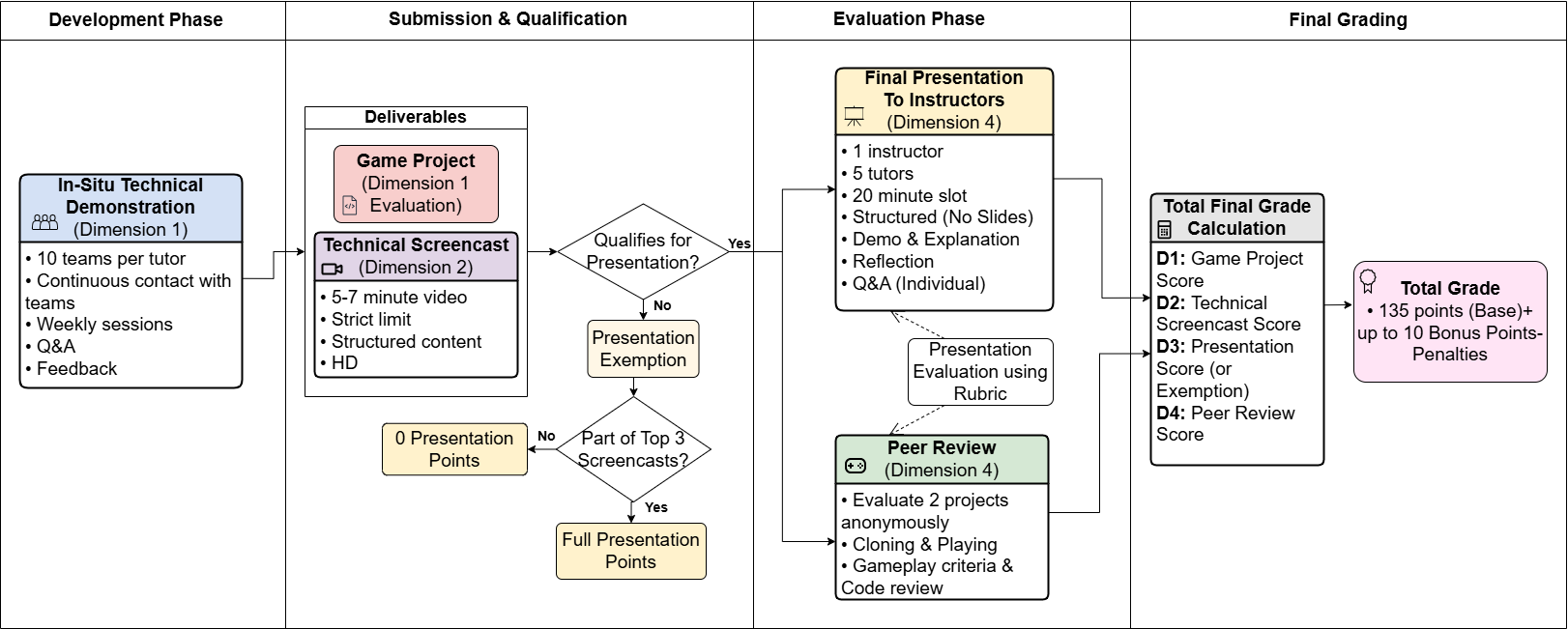}
    \caption{Multi-Modal Assessment Strategy}
    \label{Dimensions}
\end{figure*}

The assessment model developed for this course moves beyond a single high-stakes exam by distributing evaluation across four distinct dimensions. As illustrated in Figure \ref{Dimensions}, together they capture not only what students built but also how they built it, how well they can explain it, and how critically they can evaluate others' work.

\subsection{Dimension 1: In-Situ Technical Demonstration} 

It refers to the continuous technical support provided by tutors throughout the development phase. Each tutor was responsible for ten teams, exchanging contact details with them at the start of the project and remaining available for questions throughout. Teams could reach out at any point with questions, problems, or ideas, and tutors were expected to respond and help them move forward. Beyond this, the weekly tutorial sessions gave teams a weekly opportunity to sit down with their tutor, discuss progress, and get feedback on what they were building.
It is worth noting that tutors were not there to write code. Their role was to guide teams toward solutions, point out potential problems, and help them prepare for the presentation. In the end, the project was evaluated by the rubric detailed in Table \ref{tab:full_rubric}.

\begin{table}[h]
\centering
\caption{Assessment Criteria and Bonus Points for the Game Development Project}
\label{tab:full_rubric}
\begin{tabular}{|p{1.1cm}|p{4.6cm}|c|p{0.98cm}|}
\hline
\textbf{Category} & \textbf{Criteria} & \textbf{Pts} & \textbf{Bonus} \\
\hline
Game World & One entrance and multiple exits. Challenging level design with logical connections. Static traps and dynamic enemies. Keys for opening exits. & 20 & up to 3 \\
\hline
Main Character & Movement in four directions. Collision system. Multiple lives. The camera follows the character. & 15 & up to 3 \\
\hline
GUI & Game menu with start, settings, and exit options. Pause menu. Base HUD. Victory and Game Over menus. & 15 & up to 1 \\
\hline
Sound Design & Background music during gameplay and menus. Sound effects. & 10 & up to 1 \\
\hline
Graphics & Appropriate visual style on different screen sizes without discomfort or graphical issues. & 15 & up to 2 \\
\hline
Code Structure & Object-oriented implementation. No code duplication. Proper use of inheritance, delegation, and method extraction. & 15 & -- \\
\hline
Documen-tation & JavaDoc. Comments in long methods. README file detailing code structure, game use, and mechanics. & 10 & -- \\
\hline
\textbf{Total} & & \textbf{100} & \textbf{up to 10} \\
\hline
\end{tabular}
\end{table}

\subsection{Dimension 2: Technical Screencast}

The second dimension required each team to produce a technical screencast. This is a recorded video of their game, between five and seven minutes long, with penalties applied for submissions outside this range. The strict time limit required teams to be concise and selective about what they showed, which in itself is a communication skill.

The content was structured in three parts. Teams started with a brief introduction covering the game concept and storyline, then moved into a gameplay demonstration showing the core mechanics. Any bonus features the team implemented were also expected to appear here. The final part was more technical, as teams had to explain how they approached the implementation. They were also asked to reflect on challenges they faced and how they resolved them.

On the technical side, teams recorded at 1080p or higher, with clear audio narration, and submitted a link to their video hosted on a platform of their choice. The screencast was an assessment deliverable in its own right and served as a qualifier for the live presentation. Teams that did not submit a working screencast were not allowed to present. As an added motivation, the three best screencasts were exempt from the live presentation and received full presentation points automatically. The screencasts were evaluated following the rubric detailed in Table \ref{tab:screencast_rubric}.

\begin{table}[h]
\centering
\caption{Screencast Grading Rubric}
\label{tab:screencast_rubric}
\begin{tabular}{|p{1.4cm}|p{5.5cm}|c|}
\hline
\textbf{Category} & \textbf{Criteria} & \textbf{Pts} \\
\hline
Technical Explanation & 
Architectural and OOP decisions are clearly articulated. Challenges and solutions are reflected upon. & 6 \\
\hline
Gameplay Demo & 
All mandatory mechanics are shown. Bonus features, if any, are clearly distinguished from the baseline. & 6 \\
\hline
Structure \& Narration & 
Logical flow across all three parts. Narration is fluent, clear, and free of excessive fillers. & 4 \\
\hline
Introduction & 
Game concept and original storyline are presented concisely at the opening. & 2 \\
\hline
Technical Quality & 
Recorded at 1080p or higher with clear, clean audio throughout. & 2 \\
\hline
\textit{Penalty} & 
Video outside the 5--7 minute window. & $-$2 \\
\hline
\textbf{Total} & & \textbf{20} \\
\hline
\end{tabular}
\end{table}

\subsection{Dimension 3: Formal Presentation to Instructors}

The third dimension was an in-person presentation to one instructor and five tutors. Each team had a 20-minute slot: 15 minutes for the presentation itself and 5 minutes for technical setup beforehand. Being on time was a hard requirement. Teams that arrived late had less time to present, with no adjustments to grading.

The presentation followed a structure with no slides. Teams started with a brief introduction covering the game concept and story, then demonstrated and explained their features interactively, starting with the mandatory requirements before moving on to any extensions they implemented. They closed with a short reflection on the challenges they faced and what made their game unique. A Q\&A session followed, during which the panel directed questions to individual team members. All members were expected to be able to answer questions.
Everyone received the same grade based on the rubric in Table \ref{tab:presentation_rubric}, so teams had an incentive to make sure the code was shared evenly.

\begin{table}[h]
\centering
\caption{Presentation Grading Rubric}
\label{tab:presentation_rubric}
\begin{tabular}{|p{1.6cm}|p{5.5cm}|c|}
\hline
\textbf{Category} & \textbf{Criteria} & \textbf{Pts} \\
\hline
Structure & 
Logical progression and easy-to-follow flow throughout the presentation, without jumping between unrelated topics. & 5 \\
\hline
Speaker Skills & 
Clear pronunciation and voice projection. Delivered freely and fluently, without reading from the screen or using excessive fillers. & 5 \\
\hline
Introduction \& Conclusion & 
Original and engaging opening and closing. Students summarize the task in their own words rather than reading it aloud. & 3 \\
\hline
Time Management & 
Efficient use of the allocated slot. Presentation must not exceed 15 minutes. & 2 \\
\hline
\textbf{Total} & & \textbf{15} \\
\hline
\end{tabular}
\end{table}

\subsection{Dimension 4: Structured Peer-Review Process}

The fourth dimension was a structured peer-review process conducted after project submission and before the final presentations. Each team was assigned two other projects to evaluate, using the same grading requirements as the instructors used for the projects. To keep evaluations fair, neither the reviewing team nor the reviewed team could identify each other, and all submissions were handled through a dedicated form accessible only to instructors.

The review had two parts. Teams had to clone and play the assigned game, evaluating it as both players and programmers by first reviewing the gameplay criteria and then the code.

This peer review process has been studied and reported in detail in \cite{berrezueta2025coders}. This shows that first-semester students can engage meaningfully with structured peer assessment when given clear requirements and an anonymous setting, and that the process fosters evaluative and critical thinking.

\section{Implementation and Challenges}\label{IC}

\subsection{Logistical and Organizational Challenges}

The course enrolled 138 students, organized into 46 teams of 3.  Teams were self-formed and self-named, which, from the outset, encouraged a sense of ownership and identity around their project. To support them through the development process, five tutors were made available, four male and one female, each responsible for supervising between eight and ten teams on average. At the start of the project, each team selected one tutor as their primary point of contact for technical guidance, presentation 
preparation, screencast feedback, and peer-review clarifications.

Communication between teams and tutors was deliberately kept flexible. Teams could reach out through the course's online platform, engage during scheduled weekly tutorial sessions, or contact their tutor through informal digital channels. This open-ended structure lowered the barrier for asking questions and allowed tutors to adapt to the pace and needs of individual teams. When a question exceeded a tutor's expertise, the course instructor served as the final escalation point, ensuring that no team was left without a timely response.

One challenge introduced by this model was the uneven distribution of workload among tutors. Teams did not distribute uniformly, and certain tutors attracted more teams due to language or cultural proximity, as discussed further in Section~\ref{D}. Managing this imbalance required ongoing coordination among the instructional team throughout the project period.

\subsection{Training Students for Peer Review}

The peer-review process was conducted through a structured evaluation form developed collaboratively by the tutors and the course instructor. The rubric was designed to be as objective as possible, covering both gameplay criteria and code quality to minimize evaluator bias and ensure consistency across reviews. The assignment algorithm was constructed so that each team was assigned exactly two other projects to evaluate, with no reciprocal assignments, meaning that no team could be assigned to evaluate a team that was also evaluating them. No team appeared twice as either reviewer or reviewee in the same pairing set.

All reviews were submitted through a dedicated form accessible only to instructors, ensuring full anonymity for both parties. To support students during the review itself, one of the weekly tutorial sessions in the final phase of the course was designated as a supervised peer-review session. Teams attended in person and could ask tutors for clarification on technical issues encountered while running the assigned game, as well as guidance on interpreting the rubric criteria or submitting their evaluation correctly. This scaffolded setting helped mitigate the risk of poorly calibrated or incomplete reviews, particularly from students with no prior experience in formal code assessment.

\subsection{Managing Multimedia Deliverables}

Managing three distinct types of deliverables across 46 teams required a clearly defined submission pipeline for each component. The project source code was submitted through the Artemis platform, which allowed students to clone the project skeleton directly into a version-controlled repository and commit their progress iteratively up to the final deadline. This approach not only standardized the submission process but also gave tutors visibility into development activity over time.

The screencast was submitted as a hosted video link, also through Artemis, where teams uploaded the URL to a platform of their choice. To ensure technical quality, teams were required to record at a minimum resolution of 1080p with clear audio narration. Submissions outside the required five-to-seven-minute window were penalized, and teams that failed to submit a functioning screencast were not eligible to present. This hard dependency reinforced the screencast as a genuine assessment component rather than an optional extra.

The peer-review evaluations were collected through the dedicated collaborative form described above, with responses visible only to instructors and anonymity preserved throughout. Finally, the live presentations were scheduled across three consecutive days, with each team allocated a 20-minute slot in a room equipped with audiovisual infrastructure. The back-to-back scheduling required careful coordination to ensure that panel members, including the course instructor and up to 5 tutors, were present for all sessions. That room transitions between teams remained efficient.

\section{Results and Analysis}\label{R}

\subsection{Impact on Retention of Programming Fundamentals}

Of the 46 teams, 89\% submitted a project meeting all mandatory functional requirements, including collision detection, key-based exit unlocking, multi-life management, and object-oriented class hierarchies. Control structures and basic data types were present in virtually all submissions, while more advanced constructs showed greater variance: 67\% demonstrated correct inheritance hierarchies with at least three abstraction levels, and only 41\% implemented a recursive algorithm for maze traversal or enemy pathfinding.

Code structure scores averaged 9.3 out of 15 (SD = 3.1) and documentation averaged 6.8 out of 10 (SD = 2.4), confirming that functional implementation was more accessible to first-semester students than code quality and documentation.

\subsection{Communication Skills and Critical Thinking}

Screencast scores averaged 14.2 out of 20 (SD = 3.1). Gameplay demonstration and narrative coherence were the strongest segments, while technical explanation was the most common weakness. Notably, teams that scored well on the screencast consistently performed better in the live presentation, suggesting that producing the screencast served as an effective rehearsal.

Presentation soft skills scores averaged 7.6 out of 15 (SD = 1.7). Structure and speaker delivery were generally competent, while the quality of the introduction and conclusion varied most across teams. The Q\&A component exposed gaps in collective code ownership: approximately 31\% of teams struggled to answer questions that crossed individual implementation boundaries, indicating uneven workload distribution within those teams, as explained also in \cite{berrezueta2024code}.

\subsection{Student Performance Metrics and Comparison}

Table~\ref{tab:performance_summary} summarizes scores across all graded dimensions. The overall project mean was 71.7 out of 100 (SD = 12.4), ranging from 32 to 108, inclusive of bonus points. The distribution was approximately normal with a slight positive skew, confirming that the baseline was accessible to most students while the bonus structure differentiated the strongest teams. Graphics achieved the highest relative attainment (78\%), while Documentation was the lowest (68\%). Bonus points were awarded to 63\% of teams, with Game World and Main Character being the most commonly extended categories.

\begin{table}[h]
\centering
\caption{Student Performance Across Assessment Dimensions}
\label{tab:performance_summary}
\begin{tabular}{|p{3.2cm}|c|c|c|c|}
\hline
\textbf{Dimension} & \textbf{Max} & \textbf{Mean} & \textbf{SD} & \textbf{Min} \\
\hline
Game World       & 20 & 14.5 & 3.2 & 6 \\
\hline
Main Character   & 15 & 11.2 & 2.8 & 4 \\
\hline
GUI              & 15 &  7.4 & 1.9 & 2 \\
\hline
Sound Design     & 10 &  7.1 & 2.1 & 0 \\
\hline
Graphics         & 15 & 11.7 & 1.8 & 3 \\  % corrected from 7.8
\hline
Code Structure   & 15 &  9.3 & 3.1 & 3 \\
\hline
Documentation    & 10 &  6.8 & 2.4 & 1 \\
\hline
Screencast       & 20 & 14.2 & 3.1 & 6 \\
\hline
Presentation     & 15 &  7.6 & 1.7 & 3 \\
\hline
\textbf{Total}   & \textbf{135} & \textbf{93.5} & \textbf{12.4} & \textbf{32} \\  % corrected mean
\hline
\end{tabular}
\end{table}

\subsection{Student Engagement and Satisfaction}

A voluntary post-project survey collected responses from 38 teams (83\%). Overall satisfaction was high: 91\% rated the experience positively, and 84\% agreed that the game format's creative freedom increased their engagement beyond a conventional assignment. The screencast was the most novel and challenging component for 71\% of respondents, while the live presentation was the most stressful for 68\%, yet 88\% found it a valuable learning experience, and 79\% felt more confident explaining code afterward. Workload coordination was the main concern, with 76\% finding three-member collaboration more demanding than expected, and 61\% reporting scheduling pressure in the final weeks of the semester.

\subsection{Analysis of Peer-Review Quality and Reliability}

\begin{figure}[h!]
    \centering
    \includegraphics[width=1\linewidth]{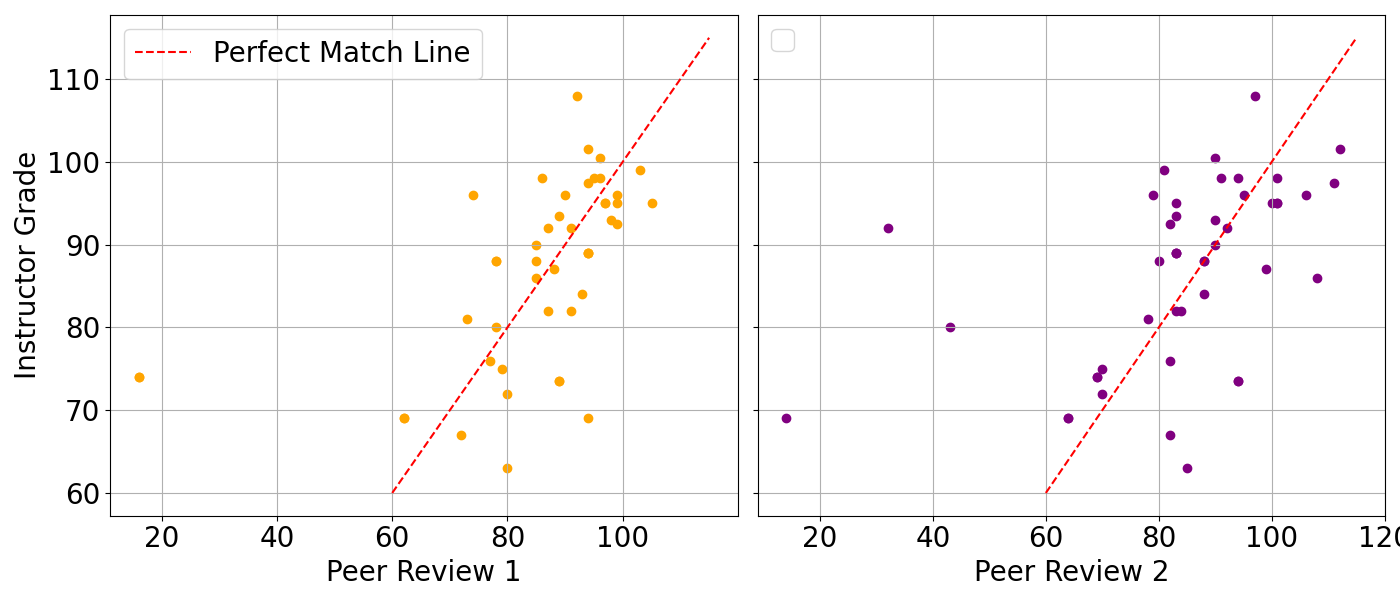}
    \caption{Comparison between peer review scores and instructor grades \cite{berrezueta2025coders}.}
    \label{Dispersion}
\end{figure}

Peer grades showed moderate correlation with instructor grades, with Pearson coefficients of 0.55 and 0.50 for the two assigned reviews, respectively \cite{berrezueta2025coders}, as illustrated in Figure \ref{Dispersion}. Agreement was higher for gameplay criteria, such as Game World and Main Character, which are directly observable during play, than for Code Structure and Documentation, which require more developed technical judgment. 

A consistent leniency bias was also observed: peer scores averaged 8.3\% higher than instructor scores, with the largest gap in Code Structure (2.1 points) and the smallest in Sound Design (0.4 points). This is a well-documented phenomenon in peer assessment \cite{che2025formative}, reinforcing the role of peer grades as a formative complement rather than a summative substitute. All teams reported believing their reviews were conducted fairly, and 83\% said they enjoyed the process \cite{berrezueta2025coders}.

\section{Discussion}\label{D}

Results reveal, in practice, how the framework can travel beyond its original context, uncover unexpected patterns in how students gravitated toward tutors, and offer an honest look at where the study falls short.

\subsection{Findings}

\note{\textbf{Finding 1:} The multimodal assessment framework was designed and evaluated within a specific institutional context, but its core structure is not inherently tied to that context.}

The four dimensions (1) continuous mentoring, (2) screencast, (3) live presentation, and (4) peer review are modular by nature, meaning each can be adopted independently or recombined to suit the constraints and goals of a given course.

\note{\textbf{Finding 2:} Any complex, open-ended project requiring design decisions, concept integration, and a tangible deliverable can serve to apply this methodology.}

The game development vehicle, while highly effective as an engagement mechanism, is not a prerequisite for the framework to function. For example, in data structures courses, a project centered on implementing and benchmarking a set of algorithms could support the same assessment dimensions. In web development or databases courses, a team-built application would provide an equally rich substrate for screencasts, peer code review, and live defense. Scalable automated tools, such as version-control analytics platforms, can further reduce the administrative overhead of tracking progress across many teams simultaneously. The key requirement is that the project is open-ended enough to generate genuine design decisions worth explaining and evaluating.

\note{\textbf{Finding 3:} Peer review is a scalable, low-overhead assessment method that generates a reliable evaluative signal even among novice students, with accuracy expected to improve as onboarding is refined.}

Peer review scales well with cohort size: the process is asynchronous and form-based, so volume increases without significant logistical overhead. The main upfront investment, rubric design, and a calibration session are a one-time cost per course iteration. The moderate correlation with instructor grades observed in this study \cite{berrezueta2025coders} confirms that even first-semester students produce a useful evaluative signal, one that is likely to improve further with better onboarding in future iterations.

Overall, the framework represents a scalable blueprint rather than a rigid prescription. Educators looking to adopt it are encouraged to begin with the dimensions most aligned with their existing infrastructure, such as adding a screencast requirement to an existing project, and expand toward the full model as institutional capacity allows.

\subsection{Observable students' patterns}

%An interesting pattern emerged in how students selected their tutors. Of the five available tutors, four were male, and one was female. Female-dominated teams showed a clear preference for the female tutor, whereas male-dominated teams were more evenly distributed among the male tutors. Similarly, students consistently gravitated toward tutors who shared their nationality or linguistic background, with Chinese, Russian-speaking, Albanian, Spanish, Arab, and Romanian students each clustering around their respective tutors, suggesting that students seek guidance from those they perceive as similar to themselves.

%While shared language lowers the barrier for asking questions and has been linked to better project outcomes through improved communication ease \cite{andree2025soft}, these clustering tendencies carried a practical cost: uneven tutor workload distribution and reduced exposure to cross-cultural collaboration, a competency of growing importance in professional software development environments.

An interesting pattern emerged in how students selected their tutors. Of the five available tutors, four were male, and one was female.
The male tutors attracted the largest share of all-male teams. One male tutor alone supervised nearly 40\% of all-male teams. The female tutor, by contrast, attracted 75\% of all-female teams, and 60\% of her overall caseload came from female-majority teams, while she received only one all-male team.

Another pattern emerged along national and linguistic lines. Students consistently gravitated toward tutors who shared their nationality or linguistic background. This mirrors the team formation patterns observed in the project itself, where approximately 46\% of the teams were nationally homogeneous at the regional level, meaning all members shared the same broad cultural or geographic origin. 

%East Asian students — predominantly Chinese — showed the strongest clustering tendency, with 62\% of East Asian teams composed entirely of East Asian students, and nine out of ten of those being all-Chinese. South Asian students followed, with 50\% of their teams being entirely South Asian. Eastern European students clustered in 43\% of cases. By contrast, MENA students, despite being a substantial group, showed lower homogeneity at 25\%. African students formed no homogeneous teams.

While shared language lowers the barrier for asking questions and has been linked to better project outcomes through improved communication ease \cite{andree2025soft}, these clustering tendencies carried a practical cost: uneven tutor workload distribution and reduced exposure to cross-cultural collaboration, a competency of growing importance in professional software development environments.

\subsection{Limitations of the Study}

While the results of this study are encouraging, several limitations must be acknowledged when interpreting the findings and considering their generalizability.

First, the study was conducted in a single course at a single institution, with a cohort of 138 students drawn predominantly from the B.Sc. in Information Engineering program. This limits the external validity of the findings, as the cultural composition of the student body, the institutional support structures, and the specific nature of the degree program may not be representative of introductory programming courses elsewhere. 

Second, the absence of a control group is a significant methodological constraint. Because all students in the course participated in the same multi-modal assessment framework, it is not possible to isolate the contribution of any single dimension. 

Finally, the study lacked a longitudinal component. It remains unknown whether the professional competencies developed during this course, including communication, peer evaluation, and technical documentation, persist into subsequent semesters or translate into measurable advantages during industry placements or advanced coursework.

\section{Conclusions and Future Work}\label{C}

This paper presented a four-dimensional assessment framework for introductory programming courses that goes beyond functional correctness to evaluate communication, critical thinking, and peer evaluation skills. Deployed across a cohort of 138 first-semester students through a 2D Maze Runner game project, the framework combined continuous tutor mentoring, a technical screencast, a live presentation, and structured peer review within a single cohesive model. Results showed that 89\% of teams met all mandatory requirements, overall project satisfaction reached 91\%, and peer-review grades correlated moderately with instructor scores, confirming that even first-semester students can engage meaningfully with formal code assessment when given clear structure and anonymity.

The framework is modular and transferable. Each assessment dimension can be adopted independently; the game project can be replaced by any open-ended artifact that elicits genuine design decisions; and the submission pipeline scales to larger cohorts with minimal overhead. The rubric design, separating functional, structural, and soft-skills criteria while rewarding creativity through bonus points, provides a ready-to-adapt template for educators looking to modernize introductory computing curricula.

Future work should address three open directions. First, replication in different institutional settings and with more diverse student populations would be necessary before broader claims can be made. Second, a controlled study comparing this framework against a traditional single-deliverable model would provide stronger causal evidence for the observed outcomes. Third, a longitudinal study \cite{harju2019exploration} tracking students across subsequent semesters would determine whether the developed competencies persist and transfer to advanced coursework or industry placements. Fourth, development and use of platforms that facilitate the peer-review process in a more attractive way than a \textit{collab} form. 

\balance
\bibliographystyle{IEEEtran}
\bibliography{references} 
\end{document}